\documentclass[aps,prb,twocolumn,showpacs,groupedaddress,superscriptaddress]{revtex4}
\usepackage{graphicx}

\begin{document}


\title{Elastic to plastic crossover below the peak effect in the vortex solid of YBa$_2$Cu$_3$O$_7$ single crystals}

\author{S. O. Valenzuela}
\altaffiliation[Present address: ]{Physics Department, Harvard
University, Cambridge, MA 02138.} \affiliation{Laboratorio de
Bajas Temperaturas, Departamento de F\'{\i}sica, Universidad de
Buenos Aires, Pabell\'on I, Ciudad Universitaria, 1428 Buenos
Aires, Argentina}

\author{B. Maiorov}
\affiliation{Centro At\'omico Bariloche and Instituto Balseiro,
Comisi\'{o}n Nacional de Energ\'{\i}a At\'{o}mica, San Carlos de
Bariloche, 8400 R\'{\i}o Negro, Argentina}

\author{E. Osquiguil}
\affiliation{Centro At\'omico Bariloche and Instituto Balseiro,
Comisi\'{o}n Nacional de Energ\'{\i}a At\'{o}mica, San Carlos de
Bariloche, 8400 R\'{\i}o Negro, Argentina}

\author{V. Bekeris}
\affiliation{Laboratorio de Bajas Temperaturas, Departamento de
F\'{\i}sica, Universidad de Buenos Aires, Pabell\'on I, Ciudad
Universitaria, 1428 Buenos Aires, Argentina}


\author{}
\affiliation{}


\date{July 5, 2001}

\begin{abstract}
We report on transport and ac susceptibility studies below the
peak effect in twinned YBa$_2$Cu$_3$O$_7$ single crystals. We find
that disorder generated at the peak effect can be partially
inhibited by forcing vortices to move with an ac driving current.
The vortex system can be additionally ordered below a well-defined
temperature where elastic interactions between vortices overcome
pinning-generated stress and a plastic to elastic crossover seems
to occur. The combined effect of these two processes results in
vortex structures with different mobilities that give place to
history effects.
\end{abstract}

\pacs{74.60. Ge, 74.60. Jg}

\maketitle

The complexity of vortex dynamics in type II superconductors
arises from the competing roles of vortex-vortex interactions
promoting order in the vortex structure (VS) and random defects in
the sample promoting disorder. Complex dynamical behavior of
vortices has been usually accompanied by the phenomenon known as
the ``peak effect" (PE), which refers to a peak feature in the
critical current density, $J_c$, as a function of temperature or
magnetic
field.\cite{sorbo,pippard,lo,kes,shobo,ling1,zies,larkin,hend} The
PE has been observed below the upper critical field, $H_{c2}$(T),
in conventional superconductors\cite{sorbo,kes,shobo,hend} and
just below the melting temperature,\cite{larkin} $T_m$, in high
$T_c$ materials such as YBa$_2$Cu$_3$O$_7$ single
crystals.\cite{ling1,zies} A qualitative explanation of the PE was
proposed forty years ago by Pippard\cite{pippard} who argued that
$J_c$ increases anomalously if the rigidity of the vortex lattice
decreases faster than the pinning interaction as the normal state
is approached. This idea was further developed by Larkin and
Ovchinnikov.\cite{lo} Nevertheless, there are still fundamental
questions regarding the underlying physics of this phenomenon,
especially in relation with the appearance of topological defects
in the VS at the PE and its connection with its dynamic response.

In this paper, we report on history effects in twinned
YBa$_2$Cu$_3$O$_7$ crystals in the vicinity of the PE as observed
in ac transport and ac susceptibility measurements. Vortex states
with different degrees of mobility have been observed in both
low,\cite{kes,shobo,hend,kupf} and high $T_c$
materials.\cite{zies,ishida,sov1,sov2} It has been frequently
argued that the difference in the mobility reflects distinct
degrees of topological order in the VS as it is driven through the
random pinning potentials: in a disordered and defective VS the
interaction with pinning centers would be more efficient than in a
more ordered lattice because of a reduction of the effective shear
modulus or of the correlation volume of the
lattice.\cite{kes,hend,sov1,sov2} In fact, changes in the degree
of order of the VS have been directly observed with neutron
scattering experiments when the VS is shaken with oscillating
currents (or fields).\cite{thor}

Having these results in mind, our experimental observations
suggest that ac currents flowing during the cooling of the sample
through the PE (either injected in a transport experiment or
induced in an ac susceptibility measurement) partially inhibit the
disordering of the VS that normally occurs at the PE. As disorder
heals, the VS is trapped in a more mobile state.

An additional contribution to the vortex lattice healing arises at
low temperatures far from the PE where our results suggest that a
plastic to elastic crossover occurs.\cite{kosh} As the sample is
cooled, the vortex lattice rigidity increases and, below a certain
threshold temperature, the vortex-vortex interactions become more
relevant relative to the pinning potentials. This further
contributes in ordering the VS, which is in a more mobile state on
warming and thereby hysteretical behavior is observed.

We will first refer to our transport data and next to our ac
susceptibility results from which identical conclusions can be
derived. Resistance measurements in crystal A\cite{crystalA}
employed the standard four probe technique with contacts made with
silver epoxy over evaporated gold pads (contact resistance below 1
$\Omega$). The sample (dimensions 0.7 mm $\times$ 0.3 mm $\times$
10 $\mu$m) has a sharp resistive zero field transition at 93.2 K
and a transition width of 500 mK (10\%-90\% criterion). Twin
boundaries (TB) observed with polarized light microscopy are
distributed as indicated in the inset of Fig.1.

A sinusoidal ac current $I_{ac}$ (peak) = 30 mA of frequency $f =
37$ Hz was applied in the $a-b$ plane of the sample, $45^{\circ}$
away from the TB planes. The voltage signal was preamplified with
a low noise transformer and its in phase value, $V(T)$, was
measured with a lock-in amplifier as a function of temperature. We
found that Joule dissipation at the current contacts drove the
sample 1.4 K above the measured temperature. The temperature shift
(assumed constant for our temperature window) occurs almost
immediately after the measuring current is applied (\textit{i.e},
at time scales well below our amplifier integrating time
constant). This shift is straightforwardly determined by matching
the {\it linear} normal state resistivity curves, $R(T)$, for
$I_{ac}$ = 30 mA with the $R(T)$ curves obtained with a much lower
applied current for which heating is negligible. We have therefore
added 1.4 K to our thermometer readings to plot our data as a
function of the temperature of the sample.

In Fig. 1, left inset, we show the zero field transition in dotted
line as compared to the transition in applied dc field shown in
full line. The magnetic field $H_{dc} =3$ kOe was applied at an
angle $\theta = 20^{\circ}$ from the crystallographic $c$ axis,
and its direction was chosen so that $H_{dc}$ pointed out of all
the twin planes to avoid the Bose-glass
phase\cite{sov1,nelson,boris} (see inset in Fig 1). The main panel
of Fig. 1 is an enlarged area of the transition shown in the upper
inset, and it presents transport measurements that correspond to
different thermomagnetic histories of the sample. Our measurements
were performed by varying  $T$ in the direction indicated by the
arrows at a rate of $\pm$ 0.3 K/min. Full symbols were obtained in
a field and current cooling procedure ($I_{ac}$C C), {\it i.e.}
the dc magnetic field and the ac current were applied in the
normal state and the sample was then cooled. Empty symbols
represent the warming curve ($I_{ac}$C W) obtained immediately
after measuring the $I_{ac}$C C curve (full symbols).

\begin{figure}[t]
\includegraphics[width=3.2in]{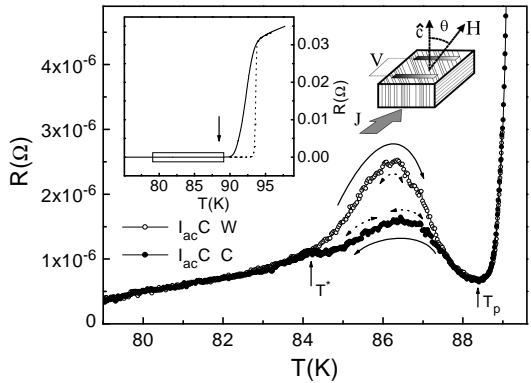} \vspace{-2mm}
\caption{Hysteretic transport measurements in a twinned YBCO
single crystal (sample A) for $H_{dc}$=3 kOe, $I_{ac}$=30 mA and
$f$=37 Hz. The inset shows the resistive transition in zero dc
field (dotted line) and with $H_{dc}$=3 kOe (full line). The
marked area is enlarged in the main panel. Full symbols were
obtained in a dc-field and current cooling procedure ($I_{ac}$C
C). Open symbols represent the warming curve ($I_{ac}$C W)
measured immediately after obtaining the $I_{ac}$C C curve.
Temperature was varied in the directions indicated by the arrows.
The crossover temperature, $T^*$, and the peak temperature, $T_p$,
are indicated by arrows (see text).} \label{fig1}
\end{figure}

When the temperature is increased, the resistance first increases,
then starts decreasing and finally increases again. Since the
resistance is related to the average vortex velocity, the dip in
the resistance indicates a less mobile VS. This behavior signals
an enhancement of pinning, and the dip in the resistance is
accompanied by a peak in $J_c$ identified as the
PE.\cite{ling1,hend} The temperature where it happens, $T_p$, is
indicated with an arrow. Above the PE, the critical current
reduces to zero at the melting line.\cite{boris,hugo}  The main
results in Fig. 1 are that, below $T_p$, we find a hysteretic
behavior that is limited to a temperature interval around the peak
in the resistance and that the response becomes reversible below a
well defined temperature $T^*\sim 84.2$ K.

It has been shown that history effects observed in ac
susceptibility in YBa$_2$Cu$_3$O$_7$ can be understood in terms of
a dynamical reordering of the VS caused by the induced ac currents
that shake the vortex system.\cite{sov1,sov2} Moreover, the ac
response changes in an approximately logarithmic way with the
number of ac cycles.\cite{sov2} In transport experiments, it is
the applied ac current that forces vortices to move inside the
sample and a dynamical reordering occurs. Non-saturated ordering
as a possible explanation for the different vortex mobilities in
the cooling and warming curves of Fig. 1 is discarded by the
observation of reversible partial temperature cycles performed as
indicated by dotted arrows in Fig. 1.\cite{com1} This result shows
that the hysteretical behavior is related to a change in the
character of the vortex lattice motion at $T^*$. This change, we
believe, is determined by the relation between the pinning induced
shear stress and the plastic limit of vortex motion associated
with the vortex lattice shear modulus $c_{66}$ (see for example Ref.
\onlinecite{kosh}). At low temperatures, well below the PE, the VS
becomes more rigid and it enters an elastic regime where elastic
interactions of the VS dominate over pinning. The enhanced elastic
interaction enables the VS to be in a more mobile state on
warming. Notice that at high temperatures, where elastic constants
have dropped, the $I_{ac}$C C and $I_{ac}$C W cases coincide.

Following these results, there are two processes that may lead to
the increase in the order and mobility of the VS: one process is
related to the increase of the elastic constants over pinning at
low temperatures whereas the second one is due to the vortex
motion induced by the driving current (at all temperatures).
Between $T^*$ and $T_p$, the VS can contain a certain density of
defects that the ac current appears not to be able to heal in our
experimental time scales.

A question that arises is what consequences has cooling the sample
through the PE without rearranging the VS (with a driving
current). To answer this question, the sample was cooled from the
normal state with applied dc field and with {\it zero applied
current} down to a certain starting temperature, $T_s$, and then
the warming curve was measured. This Z$I_{ac}$C W procedure was
repeated reaching each time a different $T_s$. The results are
plotted in full symbols in Fig. 2. For comparison, we also plot
the $I_{ac}$C W results of Fig. 1 in open symbols. These are
revealing results. The measured resistance follows different paths
as a function of temperature if $T_s$ is above a certain crossover
value ($\sim 84.2$ K). It should be noted that the sample has in
fact been cooled 1.4 K below the temperature at which each
Z$I_{ac}$C W curve in Fig. 2 seems to start (because of heating).
For measurements that begin above the crossover temperature, the
VS has less mobility and the highest measured Z$I_{ac}$C W
dissipation level that corresponds to the lower starting
temperature ($< 70$ K) is not reached. As the starting temperature
increases towards the PE temperature, $T_p$, this becomes more
evident. At high enough $T_s$'s, the ac current is unable to move
the lattice at levels that are comparable with, for example, those
obtained through the $I_{ac}$C W procedure. In contrast, at
temperatures below the crossover, the highest measured Z$I_{ac}$C
W dissipation level is reached after a rapid transient. This
transient is due to a dynamical reordering that occurs after the
measuring ac current is applied and leads to a more mobile
state.\cite{sov1}

\begin{figure}[t]
\includegraphics[width=3.2in]{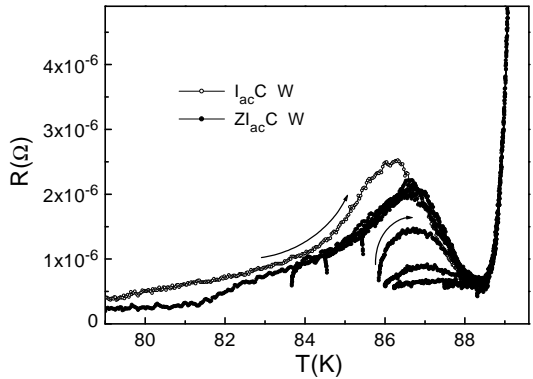} \vspace{-2mm}
\caption{Resistance {\it vs} temperature measurements obtained on
warming after cooling to different starting temperatures, $T_s$,
in zero current, Z$I_{ac}$C W (full symbols, sample A). $T_s$ =
69, 82.2, 83.1, 84.1, 84.4, 84.6, 84.8 and 86.1 K. For comparison,
it is shown the $I_{ac}$C W curve in open symbols (see Fig. 1).
$H_{dc}$ = 3 kOe.} \label{fig2}
\end{figure}

It is important to note that dissipation in the Z$I_{ac}$C W case
is lower than in the $I_{ac}$C W one at low temperatures. This
implies that an overall more ordered VS is obtained when cooling
the sample in the presence of the ac current. As predicted in Ref.
\onlinecite{sov1}, we suspect that this is due to the
inhomogeneous flow of the current that leads, in the Z$I_{ac}$C W
case, to the reordering of the VS only close to the border of the
sample where the highest currents flow. Another possible
explanation is that the current inhibits the formation of a
multiple domain (polycrystalline) VS when crossing the PE.

We turn now to our ac susceptibility results. We performed
measurements in sample B\cite{crystalB} (dimensions 1 mm $\times$
0.5 mm $\times$ 30 $\mu$m) using a conventional mutual inductance
susceptometer at a frequency $f$ = 10.22 kHz and ac magnetic field
amplitude $h_{ac}= 2$ Oe applied in the $c$ direction. This ac
field is superimposed to a constant magnetic field $H_{dc} = 3$
kOe oriented as described above. The distribution of TB 's in
crystal B is similar to that in crystal A (see inset in Fig. 1).
The sample has a sharp zero field superconducting transition at 92
K and a transition width of 300 mK (10\%-90\% criterion) as
determined with $h_{ac}=1$ Oe.

\begin{figure}[t]
\includegraphics[width=3.2in]{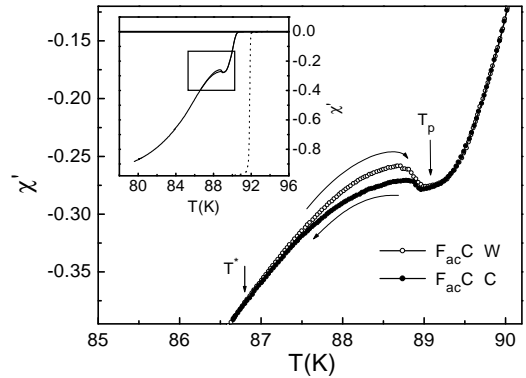} \vspace{-2mm}
\caption{Hysteretic real ac susceptibility, $\chi'(T)$, in a
twinned YBa$_2$Cu$_3$O$_7$ single crystal (sample B) for $H_{dc}$=3 kOe,
$h_{ac}$=2 Oe and $f$=10.22 kHz. The inset shows the sharp zero
field transition in dotted line and the hysteretic behavior for
$H_{dc}$=3 kOe in full line. The marked area is enlarged in the
main panel. Full symbols were obtained in a dc and ac field
cooling procedure ($F_{ac}$C C). Open symbols represent the
warming curve ($F_{ac}$C W) measured immediately after obtaining
the $F_{ac}$C C curve. Temperature was varied in the directions
indicated by the arrows. The crossover temperature $T^*$ and the
peak temperature, $T_p$, for this sample are indicated by arrows
(see text).} \label{fig3}
\end{figure}

Fig. 3 shows the real component of the ac susceptibility as a
function of temperature, $\chi$'(T). Temperature was varied as
indicated by the arrows. The upper inset shows the superconducting
zero field transition in dotted line and the $H_{dc}=3$ kOe
transition in full line. The dc field transition is enlarged in
the main panel, where hysteretic response in observed.  The lower
curve corresponds to a field cool procedure. The dc and ac
magnetic fields were applied in the normal state before cooling
the sample. We refer to this case as $F_{ac}$C C. The curve
$F_{ac}$C W was measured on warming immediately after measuring
the $F_{ac}$C C curve. In the last case, the sample clearly
presents lower shielding capability, $|\chi'|$, at intermediate
temperatures, in accordance with the higher mobility that
corresponds to a more ordered VS (compare with Fig. 1). The ac
currents induced by the applied ac field seem to have the same
effect on the VS as the ac transport currents, namely, they enable
the vortex system to access to more mobile states. The PE is also
observed with this technique, where the shielding capability
increases anomalously with temperature and a dip in $\chi$'
appears before the VS melts.

In Fig. 4 we present ac susceptibility data that match our
transport data of Fig. 2. In open symbols, we have plotted the
$F_{ac}$C W data (of Fig. 3) as a reference. Solid symbols
represent the warming curves that were measured immediately after
cooling the sample down to different temperatures in zero applied
ac field (Z$F_{ac}$C W). Note that the shielding capability in
these warming curves is always larger than in the $F_{ac}$C C or
$F_{ac}$C W procedures. As the system is cooled to different
temperatures with no applied ac field, different responses are
observed. At low temperatures, after a short transient behavior, a
response independent of the starting temperature is reached. As
the starting temperature rises and approaches $T_p$, the VS
remains less mobile, the sample presents stronger shielding
capability and the transient behavior does not lead to a common
response. It is worth pointing out that performing equivalent
analyses, we have drawn identical conclusions from the transport
and ac susceptibility measurements.

\begin{figure}[t]
\includegraphics[width=3.2in]{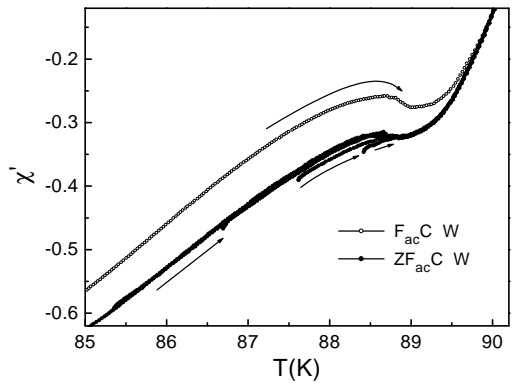} \vspace{-2mm}
\caption{$\chi'(T)$ measurements obtained on warming immediately
after cooling to different temperatures in zero ac magnetic field,
Z$F_{ac}$C W (full symbols, sample B). Temperature was varied in
the directions indicated by arrows. For comparison, it is shown
the $F_{ac}$C W in open symbols (see Fig. 3). $H_{dc}$ =3 kOe. }
\label{fig4}
\end{figure}

To summarize, we have presented ac transport and ac susceptibility
measurements in YBa$_2$Cu$_3$O$_7$ single crystals. We showed that the VS orders
assisted by the motion induced by ac currents. The increase of
elastic interactions against pinning at low temperatures leads to
a crossover in the nature of the vortex motion that favors this
reordering. This crossover from a plastic regime (which defines
the characteristic motion of vortices at temperatures close to the
peak effect) to an elastic regime at low temperatures would be
responsible for the observed hysteresis between $T^*$ and $T_p$.

The authors acknowledge F. de la Cruz and A. Silhanek for fruitful
discussions. S.O.V. thanks L. Civale for his assistance in Centro
At\'omico Bariloche. E.O. and V.B. are members of CONICET. This
research was partially supported by UBACyT TX-90, CONICET PID
N$^{o}$ 4634.


\begin{thebibliography}
\bibitem{com1} Intermediate levels of dissipation have been
observed for fast cooling or warming procedures, where the small
number of applied ac cycles per temperature interval partially
reorder the VS.\cite{sov1,sov2}
\end{thebibliography}
\end{document}